\documentclass{article}

\PassOptionsToPackage{numbers, compress}{natbib}

    \usepackage[preprint]{neurips_2022}

\usepackage[utf8]{inputenc} 
\usepackage[T1]{fontenc}    
\usepackage{hyperref}       
\usepackage{url}            
\usepackage{booktabs}       
\usepackage{amsfonts}       
\usepackage{nicefrac}       
\usepackage{microtype}      
\usepackage{xcolor}         
\usepackage{amsmath,amssymb}
\usepackage{subfigure}
\usepackage{graphicx}
\usepackage{xcolor,soul}
\usepackage[linesnumbered,ruled]{algorithm2e}
\usepackage{footnote}
\usepackage{amsthm}
\usepackage{accents}
\usepackage{bbold}
\usepackage{mathtools}
\usepackage{verbatim}
\usepackage{tikz}
\usepackage{wrapfig}

\title{Dynamic Calibration of Nonlinear Sensors with Time-Drifts and Delays by Bayesian Inference}

\author{
  Soumyabrata Talukder
    \\
  Iowa State University\\
  2520 Osborne Drive\\
  Ames, IA 50010 \\
  \texttt{talukder@iastate.edu} \\
    \And
   Souvik Kundu \\
   Iowa State University \\
   2520 Osborne Drive \\
   Ames, IA 50010\\
   \texttt{souvik@iastate.edu} \\
   \And
   Ratnesh Kumar \\
   Iowa State University \\
   2520 Osborne Drive \\
   Ames, IA 50010\\
   \texttt{rkumar@iastate.edu} \\
}

\begin{document}

\maketitle

\begin{abstract}
Most sensor calibrations rely on the linearity and steadiness of their response characteristics, but practical sensors are nonlinear, and their response drifts with time, restricting their choices for adoption. To broaden the realm of sensors to allow nonlinearity and time-drift in the underlying dynamics, a Bayesian inference-based nonlinear, non-causal dynamic calibration method is introduced, where the sensed value is estimated as a posterior conditional mean given a finite-length sequence of the sensor measurements and the elapsed time. Additionally, an algorithm is proposed to adjust an already learned calibration map online whenever new data arrives. The effectiveness of the proposed method is validated on continuous-glucose-monitoring (CGM) data from an alive rat equipped with an in-house optical glucose sensor. To allow flexibility in choice, the validation is also performed on a synthetic blood glucose level (BGL) dataset generated using FDA-approved virtual diabetic patient models together with an illustrative CGM sensor model.
\end{abstract}

\section{Introduction}
The selection of a practical sensing system typically requires the linearity and steadiness of its response because of the lack of reliable dynamic calibration techniques. For this reason, sensor response steadiness analysis has been explored in literature \cite{drobek2016mof, neshkova1996piezoelectric, wencel2018optical, tsujita2004dynamic}, and typically a sensor, even if it is precise, gets discarded whenever
an appropriate method of nonlinear time-varying compensation for its drifting characteristics is unavailable. The challenge is particularly prominent for \emph{continuous} monitoring (CM) applications \cite{vettoretti2015online, guerra2012enhancing} where a sensor drift mandates frequent recalibration, impacting the continuity of monitoring. It is thus significant to devise a general method to correct the drifting characteristics of a sensor, thereby offering added flexibility in the choice of sensors and extending their lifetime of usage. Here we refer to such a method as ``dynamic calibration'', where not just the current sensor response but its finite sequence is used for dynamically adjusting the calibration. 

The fundamental challenge in performing dynamic calibration of sensors in practical applications is twofold. First, the temporal drift of the sensing system characteristics may be nonlinear, so manually finding an appropriate form of the calibration map may be difficult in general. 
Second, the underlying dynamics of the sensing system often causes transient drift of sensitivity \cite{guerra2012enhancing, vettoretti2015online, acciaroli2017reduction} that limits the sensor from being used.  This motivates us to develop a systematic method to enable effective dynamic calibration of sensors that are nonlinear and incur drift. Our method is also data-driven, benefiting from the recent advances in ``internet of things'' and smart sensing devices, that provide a large pool of historical sensor data in many applications \cite{atzori2010internet}, offering the opportunity for their dynamic calibration and thereby increase the accuracy of their real-time readings.  

\paragraph{Related Works} Several methods exist in the literature that aims to address the problem of dynamic calibration for continuous glucose monitoring (CGM) from interstitial glucose (IG) concentrations, where the measurement accuracy is adversely affected by the underlying blood glucose (BG) to IG kinematics \cite{diabetes2006evaluation, basu2013time}. In \cite{aussedat1997user}, the authors propose calibration of the CGM sensors only at the steady states of the kinematics, which forbids the calibrated sensor to be used during the transient phase.  
In \cite{guerra2012enhancing}, the BG to IG kinematics is assumed to possess an \emph{empirical linear first-order} dynamic model, which is limiting, and where the kinematics induced distortion is proposed to be compensated by a regularized deconvolution procedure relying on a linear regression model. In \cite{vettoretti2015online}, noting the issue of temporal drift of sensitivity, the calibration map of \cite{guerra2012enhancing} is enhanced by introducing explicit \emph{linear} time-dependence, which limits its applicability. The parameters of the calibration map, owing to its linearity, are learned by regularized least squares. A time-varying calibration map is explored in \cite{acciaroli2017reduction} by making it linearly dependent on certain known time-dependent functions, which again is restrictive, and where the bilinear parameters of the map are estimated by a nonconvex iterative search employing Nelder-Mead algorithm \cite{de1997nonparametric}. In \cite{aussedat1997user, guerra2012enhancing, vettoretti2015online, acciaroli2017reduction}, the ground-truths for calibration are assumed to be \emph{noise-free}, which also is restrictive. In a later work \cite{vettoretti2019development}, a 3-component CGM error model is proposed, where the components respectively correspond to a first-order
linear dynamic model describing the BG to IG kinetics, a second-order polynomial model for the calibration error, and an autoregressive model to account for the measurement noise. Further, the dynamic calibration methods proposed in \cite{aussedat1997user, guerra2012enhancing, vettoretti2015online, acciaroli2017reduction, vettoretti2019development} require that the underlying sensing system dynamics must be linear and known (to be able to employ deconvolution), which limits their general applicability. Our method works with nonlinear time-varying systems whose dynamics are not known a priori.
 
Among others, in \cite{geng2015gaussian}, a calibration method for an environment-induced drift in a sensor is proposed by way of including the observable dynamic environmental features (e.g., temperature, pressure, etc.) as the \emph{exogenous} input to the calibration map, so that the map itself remains free from explicit time-dependence. However, every environmental feature that causes drift in the sensor's sensitivity may not be known, and/or the accurate measurement of all the known ones may not be feasible in every application, which limits the applicability of the method in general. 
A method is introduced in \cite{7927400} to account for temporal drift in wireless sensor networks (SNs) by projecting the drifted data to a feature-space and using a convolutional neural network to estimate the sensed value from those features. The drifting issue in SNs is addressed in \cite{tsujita2004dynamic} for gas sensing applications, where the rate of time-drift of each mobile sensor is assumed to be fixed, independent, and uniformly random over a given range. The true gas concentration in a neighborhood is then determined as the arithmetic mean of the values estimated by each local sensor, respectively, and each mobile sensor is recalibrated online whenever it is in the vicinity of any ground-truth value. However, the methods of \cite{7927400,tsujita2004dynamic} cannot be applied to handle the drifting sensitivity of \emph{standalone} sensors. 

\paragraph{Contributions}Our main contributions are as follows: (i) A Bayesing-inference based nonlinear non-causal dynamic calibration method for an underlying  nonlinear, time-varying, and unknown dynamics of the sensing system has been proposed, where the sensed value is estimated as its posterior conditional mean given a finite sequence of the sensor measurements and the elapsed time, starting with a Gaussian process prior, (ii) An iterative scheme is proposed for online update of the dynamic calibration whenever new data arrives during the deployment phase, (iii) The effectiveness of the method for calibration based on real measurements is validated by conducting a CGM experiment on an alive rat equipped with an in-house designed optical glucose sensor, (iv) To allow the flexibility of data, the proposed method is further validated on a synthetic blood glucose level (BGL) dataset generated using FDA-approved virtual diabetic patient models of varied age groups, together with an illustrative continuous-glucose-monitoring (CGM) sensor model. 

Gaussian process model has been used previously to identify ``forward dynamics" in the form of stochastic models \cite{kocijan2005dynamic, geng2015gaussian, eleftheriadis2017identification}, inferring the distribution of the outputs given the past sequence of inputs. In contrast, the dynamic sensor calibration is an ``inverse problem" that has not gained much attention, in which we infer an input (i.e., the sensed quantity) given a sequence of outputs (i.e., the sensor measurements). The inverse mapping in general is noncausal and poses additional challenges in our case as it involves time-drifts in sensor response, random time-varying delays, and unobservability of the state-space. Since a GP model is primarily correlation-based, there is no a priori guarantee of accurate prediction of such noncausal inverse mapping. Our result for the first time demonstrates that a GP model can perform favorably in the dynamic calibration problem of interest, as shown for the real-world optical sensor-based CGM application. 

Also unlike \cite{aussedat1997user, guerra2012enhancing, vettoretti2015online, acciaroli2017reduction, vettoretti2019development}, we do not require knowledge of the dynamics of the underlying sensing system or its linearity. Further, in contrast to \cite{geng2015gaussian}, our method does not depend on the availability of measurements of certain exogenous variables, and unlike the methods in \cite{7927400,tsujita2004dynamic}, we do not require measurements from multiple sensors for dynamic calibration.    

\paragraph{Notations}
For $\mathbf{x}\in\mathbb{R}^n$, 
$\|\mathbf{x}\|$ denotes
its Euclidean norm, and for any square matrix $\mathbf{X}\in\mathbb{R}^{n\times n}$, $|\mathbf{X}|$ and $\text{tr}(\mathbf{X})$ denote its determinant and trace respectively. 
$\mathbf{1}_n$ denotes the $n$-dimensional unit vector.

\section{Dynamic Sensor Calibration Formulation}
A sensing system is viewed as an unknown nonlinear time-varying discretized dynamical system of the following form:
\begin{equation}\label{ideal_model}
          \mathbf{x}_{k+1} = f(\mathbf{x}_k, u_k, t_k),\quad
        y_k = h(\mathbf{x}_k, u_k, t_k), \quad0\leq t_k\leq T,
 \end{equation}
where $\mathbf{x}\in\mathbb{R}^n$, $u\in\mathbb{R}$, $y\in\mathbb{R}$, and $t\in\mathbb{R}_{\geq0}$ respectively denote the latent state vector, the sensed quantity that acts as the input to the system, the sensor measured signal, and the time elapsed after the sensor deployment; their values at the $k^{th}$ discrete sample instant ($k\geq 0$) are denoted by introducing the subscript $k$.  
The explicit dependence on $t_k$ of the sensing system's state dynamics $f(\cdot, \cdot, \cdot)$ and output map $h(\cdot, \cdot,\cdot)$, captures the time drift of the system. The lifespan of the sensing element is denoted $T$. For CM applications, a sensor is replaced with a new identical one at the end of its life, and the sensing system dynamics (\ref{ideal_model}) is reinitialized with $x_k=0$, $t_k=0$.

Note in certain applications, due to restricted direct access to the sensed quantity, it may be of interest to estimate its value from another accessible variable related to the sensed quantity through unknown dynamics. Such dynamics can easily be subsumed in $f(\cdot,\cdot,\cdot)$ even though it is not an intrinsic part of the engineered sensing system. For example, the unknown BG to IG kinematics that distorts the accuracy of BGL estimation from IG level \cite{vettoretti2015online, guerra2012enhancing, aussedat1997user, acciaroli2017reduction}, can be viewed as subsumed within (\ref{ideal_model}). 

The goal of the dynamic calibration, in general, is to estimate the sensed signal values $u_k$'s in the form of $\hat{u}_k$'s from a finite sequence of associated sensor measurements $y_k$'s. Such calibration requires data concerning time-series of sensor-measured $y_k$'s together with the corresponding reference values $\tilde{u}_k$'s, recorded using an independent standard instrument and acting as ``ground truth''. For our proposed method, $\tilde{u}_k$'s are allowed to be corrupted by Gaussian noise.

\section{Proposed Dynamical Calibration for Time-Varying and Nonlinear Sensor}
Consider an unknown time-varying non-causal map relating the $k$th time-step estimate  $\hat u_k$ of the actual signal to a finite-length vector of sensor measurements $\overline{\mathbf{z}}_k:=[\mathbf{y}_{k,p,\ell}^T \quad t_k]^T$, where $\mathbf{y}_{k,p,\ell}\in\mathbb{R}^p$ stacks the past $p$ and future $\ell$ measurements $y_{k-p},~ \ldots, ~y_{k},~y_{k+1},~\ldots,~y_{k+\ell}$ in order, namely,
\begin{equation}\label{true_bc}
    \hat u_k = g(\overline{\mathbf{z}}_k),\quad\mbox{where  }\overline{\mathbf{z}}_k:=[\mathbf{y}_{k,p,\ell}^T \quad t_k]^T\mbox{ and } k\geq p; p,\ell\geq 0.
\end{equation}

Gaussian process-based supervised learning is well-known for its ``universal consistency'' and effectiveness in limited size of training dataset \cite{seeger2004gaussian}, often encountered in practice such as biomedical applications. Hence to accomplish our above goal of dynamic calibration, instead of learning a deterministic version of $g(\cdot)$, we aim to learn a statistical version of it from a Gaussian process prior, using the dataset $\mathcal{D}:=\{(\overline{\mathbf{z}}_k,\tilde{u}_k)~|~k\in\{1,\ldots,N\}\}$, which is a collection of $N$ $(\overline{\mathbf{z}}_k,\tilde{u}_k)$-pairs extracted from the available time-series of $y_k$'s and their corresponding noisy reference values $\tilde{u}_k$'s measured for calibration purposes by an independent instrument with an unknown noise variance $\sigma^2_{\tilde u}$. 
The choice of learning a statistical version of $g(\cdot)$ also enables an iterative online update of the learned map utilizing the prediction statistics. The learned statistical version of $g(\cdot)$ is referred as the ``statistical dynamic calibration map'' (SDCM), that for any given $\overline{\mathbf{z}}_k$ provides an estimate of the mean and the variance of the corresponding $\hat u_k$. 

In what follows, section 3.1 presents a Bayesian framework to learn an SDCM, and in section 3.2, we propose an iterative method for online update of an already learned SDCM when new data samples become available. 

\subsection{Statistical Model Learning}\label{stat_model} 
Let $\overline{\mathbf{Z}}:=\{\overline{\mathbf{z}}_1, \ldots,\overline{\mathbf{z}}_N\}$ denote the set of all $\overline{\mathbf{z}}$'s in the calibration dataset $\mathcal{D}$. We assume that the samples of reference values $\tilde u_k$'s form a stationary Gaussian process \cite{kac1947explicit}, where recall the unknown variance of the additive white noise present in the reference measurements $\tilde{u}$ is denoted $\sigma_{\tilde u}^2$. The prior correlation between any two $\tilde{u}$ samples is assumed to take the form of a radial basis function with two other unknown parameters, $\delta$ and $\sigma$, to be learned from the data. Formally, it follows that at any measurement instant $k\geq p$, the noisy reference values $\tilde{u}_1,\ldots,\tilde{u}_N\in\mathcal{D}$ and the $k$th instant estimator $\hat{u}_k$, are jointly Gaussian with the prior:
    \begin{equation}\label{stat_model}
        \begin{split}
            & \begin{bmatrix}
                \tilde{u}_1\\
                \vdots\\
                \tilde{u}_N\\
                \hat{u}_k
                \end{bmatrix}  
                \sim\mathcal{N}\Bigg(\mu.\mathbf{1}_{N+1}, 
                \underbrace{\begin{bmatrix}
                \underbrace{\mathbf{\Sigma}_{u}(\overline{\mathbf{Z}}, \delta, \sigma) + \sigma_{\tilde{u}}^2.\mathbf{I}_N}_{:=\mathbf{\Sigma}_{11}(\overline{\mathbf{Z}} ,\mathbf{\theta})} & \mathbf{\Sigma}_{12}(\overline{\mathbf{Z}}, \delta, \sigma,\overline{\mathbf{z}}_k)\\
                \mathbf{\Sigma}_{12}(\overline{\mathbf{Z}}, \delta, \sigma, \overline{\mathbf{z}}_k)^T & \sigma^2\\
            \end{bmatrix}}_{:=\mathbf{\Sigma}(\overline{\mathbf{Z}},\overline{\mathbf{z}}_k,\mathbf{\theta})}
            \Bigg),
        \end{split}
    \end{equation}
where to account for the effect of the white noise present in $\tilde{u}_i$'s, we have added the constant noise term $\sigma_{\tilde{u}}^2.\mathbf{I}_N$ to $\mathbf{\Sigma}_{u}$ in (\ref{stat_model}), $\mu\in\mathbb{R}$ is defined as $\mu:=\sum_{i\in\{1,\ldots,N\}}\tilde{u}_i/N$,  $\overline{\mathbf{z}}_k$ denotes the latest vector obtained by stacking up the measurements and the elapsed time (as in (\ref{true_bc})), and the $(i,j)^{th}$ element of $\mathbf{\Sigma}_{u}\in\mathbb{R}^{N\times N}$, denoted $\mathbf{\Sigma}_{u}^{i,j}$, and the $i^{th}$ element of $\mathbf{\Sigma}_{12}\in\mathbb{R}^{N}$, denoted
$\mathbf{\Sigma}_{12}^{i}$, are defined using a radial basis function kernel $\kappa(\cdot,\cdot)$ as follows:
 \begin{equation}\label{stat_model+}
\mathbf{\Sigma}_{u}^{i,j}:=\kappa(\overline{\mathbf{z}}_i, \overline{\mathbf{z}}_j):=\sigma^2.e^{-\dfrac{\|\overline{\mathbf{z}}_i-\overline{\mathbf{z}}_j\|^2}{2\delta^2}}; \quad\mathbf{\Sigma}_{12}^{i}:=\kappa(\overline{\mathbf{z}}_i, \overline{\mathbf{z}}_k):=\sigma^2.e^{-\dfrac{\|\overline{\mathbf{z}}_i-\overline{\mathbf{z}}_k\|^2}{2\delta^2}}.
    \end{equation}
Note the above chosen prior enforces the following property: A pair $(\tilde{u}_i,\tilde{u}_j)$ or $(\tilde{u}_i,\hat{u}_k)$ is deemed strongly (resp., weakly) correlated if the corresponding $\overline{\mathbf{z}}$ values are close (resp., far apart) in the euclidean distance sense, and the degree of that correlation is parameterized by $\delta$ and $\sigma$, where $\delta$ defines the exponential rate of change of correlation with respect to the said euclidean distance, and $\sigma$ scales the correlation uniformly. By virtue of the chosen kernel $\kappa(\cdot, \cdot)$, the matrix $\mathbf{\Sigma}\in\mathbb{R}^{(N+1)\times(N+1)}$ is positive semidefinite, and hence it can serve as the covariance matrix of the above multivariate Gaussian distribution.  

\subsubsection{Bayesian Inference of True Sensed Value}
Let $\mathbf{\theta}:=[\delta~~\sigma~~\sigma_{\tilde{u}}]^T$ denote the parameters to be learned from the data. In what follows, we propose a Bayesian method to estimate the sensed value $\hat{u}_k$ as a function of $\overline{\mathbf{z}}_k$ together with the parameter $\theta$, followed by a discussion about a method to find the optimal value of $\theta$ using empirical Bayes. 
Given the calibration dataset $\mathcal{D}$ and the estimator parameter $\mathbf{\theta}$, the posterior distribution of $\hat{u}_k$ can be estimated as a function of $\overline{\mathbf{z}}_k$:
    \begin{equation}
        \hat{u}_k\sim\mathcal{N}(\mu_{\hat{u}}(\overline{\mathbf{z}}_k), \sigma_{\hat{u}}^2(\overline{\mathbf{z}}_k)~|~\mathcal{D},\mathbf{\theta}),
    \end{equation}
    where $\mu_{\hat{u}}: \mathbb{R}^{p+2}\rightarrow\mathbb{R}$, $\sigma_{\hat{u}}^2: \mathbb{R}^{p+2}\rightarrow\mathbb{R}$ denote the posterior mean and variance of $\hat{u}_k$ respectively. Based on the Bayesian theory of conditionals of a multivariate Gaussian (see Theorem 4.3.1 of \cite{murphy2012machine}), the posterior mean and variance of $\hat{u}_k$ given $\mathcal{D}$ and $\theta$ are computed as follows:
    \begin{equation}\label{predict}
        \begin{split}
            \mu_{\hat{u}}(\overline{\mathbf{z}}_k)  = \mathbf{\Sigma}_{12}^T(\overline{\mathbf{z}}_k).\mathbf{\Sigma}_{11}^{-1}.(\tilde{\mathbf{u}}-\mu) + \mu,\qquad
            \sigma^2_{\hat{u}}(\overline{\mathbf{z}}_k)  = \sigma^2 - \mathbf{\Sigma}_{12}(\overline{\mathbf{z}}_k)^T.\mathbf{\Sigma}_{11}^{-1}.\mathbf{\Sigma}_{12}(\overline{\mathbf{z}}_k),
        \end{split}
    \end{equation}
    where $\tilde{\mathbf{u}}:=[\tilde{u}_1~~\ldots~~\tilde{u}_N]^T$, and note that $\mathbf{\Sigma}_{12}(\overline{\mathbf{Z}},\delta,\sigma,\overline{\mathbf{z}}_k~|~\mathcal{D}, \mathbf{\theta})\equiv\mathbf{\Sigma}_{12}(\overline{\mathbf{z}}_k)$.
    
At each sample instant during the deployment phase, 
$\mu_{\hat{u}}(\overline{\mathbf{z}}_k)$ and $\sigma^2_{\hat{u}}(\overline{\mathbf{z}}_k)$ values from (\ref{predict}) provide the expected sensed value and its variance, respectively, and the two together form the SDCM. 
Note that $\gamma(\overline{\mathbf{z}}_k):=1/\sqrt{\sigma_{\hat{u}}^2(\overline{\mathbf{z}}_k)}$ naturally serves as a measure of the confidence in the estimate $\mu_{\hat{u}}(\overline{\mathbf{z}}_k)$. This measure is used in the next section to implement an iterative online update scheme.
    
\subsubsection{Model Parameter Estimation by Empirical Bayes} Before the deployment of the sensing system, the parameter $\mathbf{\theta}$ of the prior distribution in (\ref{stat_model}) needs to be tuned. Employing empirical Bayes \cite{casella1985introduction}, we compute an optimal estimate of $\mathbf{\theta}$ that maximizes the likelihood of dataset $\mathcal{D}$. 
Since the elements of $\tilde{\mathbf{u}}$ can be viewed to be jointly sampled from $\mathcal{N}(\mu.\mathbf{1}_N,\mathbf{\Sigma}_{11}(\overline{\mathbf{Z}},\theta))$ (equivalently, $\tilde{\mathbf{u}}-\mu.\mathbf{1}_N\sim\mathcal{N}(\mathbf{0}_N,\mathbf{\Sigma}_{11}(\overline{\mathbf{Z}},\theta))$), the natural logarithm of the likelihood of $\mathcal{D}$ can be written as: 
    \begin{equation}\label{likelihood}
        \log p(\tilde{\mathbf{u}}~|~\overline{\mathbf{Z}}) =-\frac{1}{2}\big(\tilde{\mathbf{u}}^T.\mathbf{\Sigma}_{11}^{-1}(\theta).\tilde{\mathbf{u}} + \log|\mathbf{\Sigma}_{11}(\theta)|+N.\log2\pi\big),
    \end{equation}
    where note that $\mathbf{\Sigma}_{11}(\overline{\mathbf{Z}},\theta~|~\mathcal{D})\equiv\mathbf{\Sigma}_{11}(\theta)$. Then the optimal value $\theta^*$ is computed as:
    \begin{equation}\label{max_likelihood}
        \theta^*=\underset{\theta}{\mathrm{argmax}}~\log p(\tilde{\mathbf{u}}~|~\overline{\mathbf{Z}}).
    \end{equation}
The optimum is explored via gradient descent.
Due to the nonconvexity of the optimization in (\ref{max_likelihood}), gradient descent finds local optima, in general. (\ref{max_likelihood}) is solved multiple times with $\theta$ initialized at different points. 
The best among these solutions is chosen for the estimator in (\ref{stat_model}). 

\subsection{Iterative Online Update of Dynamic Calibration} \label{iterative}
The purpose of iterative online update of the SDCM is to  account for any unexpected characteristic changes of the sensing system (e.g., replacement of a component with a new one, etc.) or to fine-tune the learned SDCM for a specific application environment. 
This is particularly important in CM applications with periodic replacement of sensing elements. In such applications, often sparse measurements are obtained using a standard instrument to inspect the performance quality of the CM sensor, which can be utilized to create new application-specific calibration-data samples. Imputing such new samples, next, we propose an algorithm to iteratively update the calibration dataset $\mathcal{D}$ online and to recompute the SDCM, keeping the total number of samples $N$ fixed. In practice, $N$ will be limited to a few thousands, as it affects the complexity of SDCM computation in cubic order.

Let $(\overline{\mathbf{z}}_{new},\tilde{u}_{new})$ denote a newly observed sample. The questions we attempt to answer are: (i) whether $(\overline{\mathbf{z}}_{new},\tilde{u}_{new})$ is a sample worth adding to $\mathcal{D}$, and (ii) if so, which sample in $\mathcal{D}$ should it replace so that the total sample count remains unchanged. To answer the first question, we check if $(\overline{\mathbf{z}}_{new},\tilde{u}_{new})$ is an outlier by employing a tunable parameter $\epsilon_u\in\mathbb{R}_{\geq0}$ and deeming $(\overline{\mathbf{z}}_{new},\tilde{u}_{new})$ to be an outlier if:
\begin{equation}\label{outlier}
    \big\|\mu_{\hat{u}}(\overline{\mathbf{z}}_{new}) - \tilde{u}_{new}\big\| > \epsilon_{u}.
\end{equation}
If $(\overline{\mathbf{z}}_{new},\tilde{u}_{new})$ is an outlier, we reject it. To answer the second question, we first define a similarity matrix $\mathbf{S}\in\mathbb{R}^{N\times N}$ for the samples in dataset $\mathcal{D}$, the $(i,j)^{th}$ element of which is given by:
\begin{equation}
    \mathbf{S}^{i,j}:=e^{-c\|\mathbf{v}_i-\mathbf{v}_j\|};~~ \mathbf{v}_i:=\big[(\overline{\mathbf{z}}_i)^T~~ \tilde{u}_i\big]^T;~~i,j\in\{1,\ldots,N\};~~c\in\mathbb{R}_{\geq0}.
\end{equation}
Note $\mathbf{S}^{i,j}\in(0,1]$ serves as a measure of the similarity between the $i^{th}$ and the $j^{th}$ samples of $\mathcal{D}$, which exponentially decays with increasing $\|\mathbf{v}_i-\mathbf{v}_j\|$ at rate $c$, starting at unity when $\mathbf{v}_i=\mathbf{v}_j$. Also, note that $\mathbf{S}$ is symmetric as desired. If (\ref{outlier}) is falsified (i.e., not an outlier), then we replace the existing sample $(\overline{\mathbf{z}}_{i^*},\tilde{u}_{i^*})$ of $\mathcal{D}$ with $(\overline{\mathbf{z}}_{new},\tilde{u}_{new})$, where the index $i^*$ is decided as follows:
\begin{equation}\label{replace}
    i^*=
    \begin{cases}
        \underset{i}{\mathrm{argmin}}~\|\mathbf{v}_i - \mathbf{v}_{new}\| & \text{if}~\gamma(\overline{\mathbf{z}}_{new}) \geq \epsilon_{\gamma} \\
        \underset{i}{\mathrm{argmax}}~\sum_j\mathbf{S}^{i,j} & \text{if}~\gamma(\overline{\mathbf{z}}_{new}) < \epsilon_{\gamma}
    \end{cases},
\end{equation}
where $\epsilon_{\gamma}\in\mathbb{R}_{\geq0}$ is another tunable parameter. The rationale behind (\ref{replace}) is as follows:  If the model's inference has high confidence (i.e., when $\gamma(\overline{\mathbf{z}}^{new}) \geq \epsilon_{\gamma}$), $(\hat{\mathbf{z}}_{new},\tilde{u}_{new})$ replaces the existing sample that is closest to it; otherwise, if the model infers the current sensed value with low confidence (so there is likely no existing sample in $\mathcal{D}$ with $\overline{\mathbf{z}}_i$ close to $\overline{\mathbf{z}}_{new}$), the new sample $(\hat{\mathbf{z}}_{new},\tilde{u}_{new})$ replaces the most redundant sample (i.e., the one with the highest corresponding $\sum_j\mathbf{S}^{i,j}$ value). The overall algorithm is laid out in Algorithm \ref{algo1}, where $\mathcal{D}^k$ denotes the updated calibration dataset at the $k^{th}$ online calibration sample instant.

\begin{algorithm}\label{algo1}
\SetAlgoLined
\textbf{Input}: Initial calibration dataset $\mathcal{D}$ with $N$ samples; parameters $c, \epsilon_u,\epsilon_\gamma\in\mathbb{R}_{\geq0}$.\\
\textbf{Initialize}: $k=0$; $\mathcal{D}^k=\mathcal{D}$. \\
 \While{true}{
 $k\leftarrow k+1$.\\
 Observe new sample $(\overline{\mathbf{z}}_{new},\tilde{u}_{new})$.\\
 \eIf{(\ref{outlier}) $\text{is falsified}$}
 {
 Decide the sample index $i^*$ in $\mathcal{D}^{k-1}$ using (\ref{replace}).\\
 $\mathcal{D}^{k}=\big(\mathcal{D}^{k-1}~\bigcup~(\overline{\mathbf{z}}_{new},\tilde{u}_{new})\big)\setminus\{(\overline{\mathbf{z}}_{i^*},\tilde{u}_{i^*})\}$.\\
 Recompute the forms: $\mu_{\hat{u}}(\cdot)$ and $\sigma^2_{\hat{u}}(\cdot)$ given $\mathcal{D}^k$ to infer the true BBC values.}
 {
 $\mathcal{D}^{k}=\mathcal{D}^{k-1}$, and also alert to launch an outlier investigation process.
}
}
\caption{Recalibration by iterative update of calibration dataset}
\end{algorithm}

\section{Calibration Validation Results}
In this section, the effectiveness of the proposed dynamic calibration method is validated on synthetic CGM data as well as real CGM data collected from alive rats in a laboratory setting. The synthetic case is introduced since it offers the flexibility to choose the sensor model and varied noise levels of the reference as appropriate for effective assessment and illustration of the method's capabilities. The real dataset is from our in-house sensor on alive rat subjects. The obtained results corroborate the applicability of the proposed dynamic callibration method for calibration of a real sensor in a real-life measurement setting. 
 
\subsection{Calibration of Synthetic Continuous Glucose Monitoring Data}\label{synthetic}
Simglucose \cite{xie2018simglucose}, a Python tool used in literature for CGM-related experiments \cite{huang2021off, eichstadt2016challenges} that implements
   \begin{wrapfigure}{r}{0.7\textwidth}
    \centering
    \subfigure[BGL corrupted by 55 dB SNR white noise]{
    \includegraphics[width=0.32\textwidth]{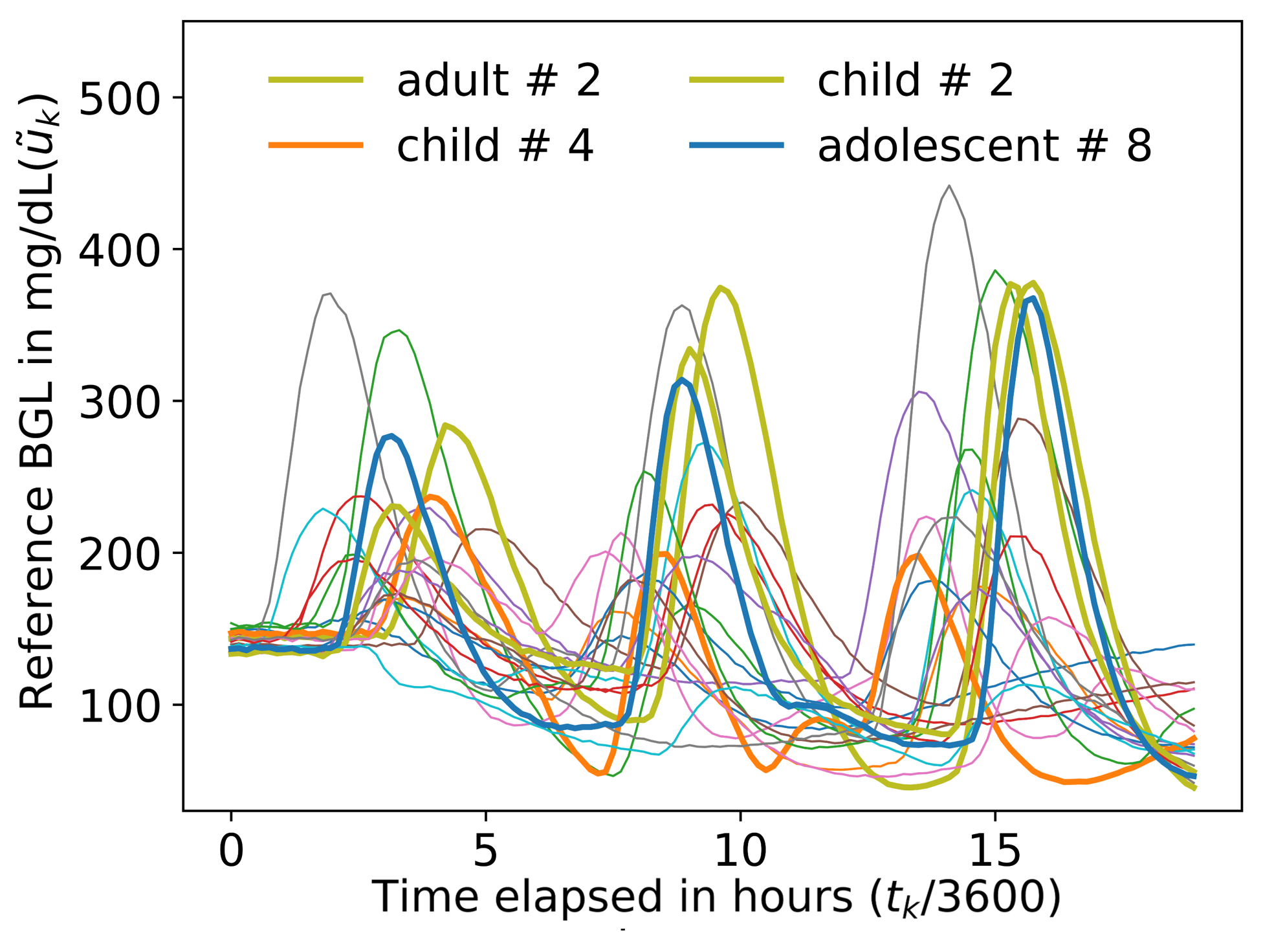}
    \label{fig:figure1}}
    \hfill
    \subfigure[Sensor measurement]{
    \includegraphics[width=0.32\textwidth]{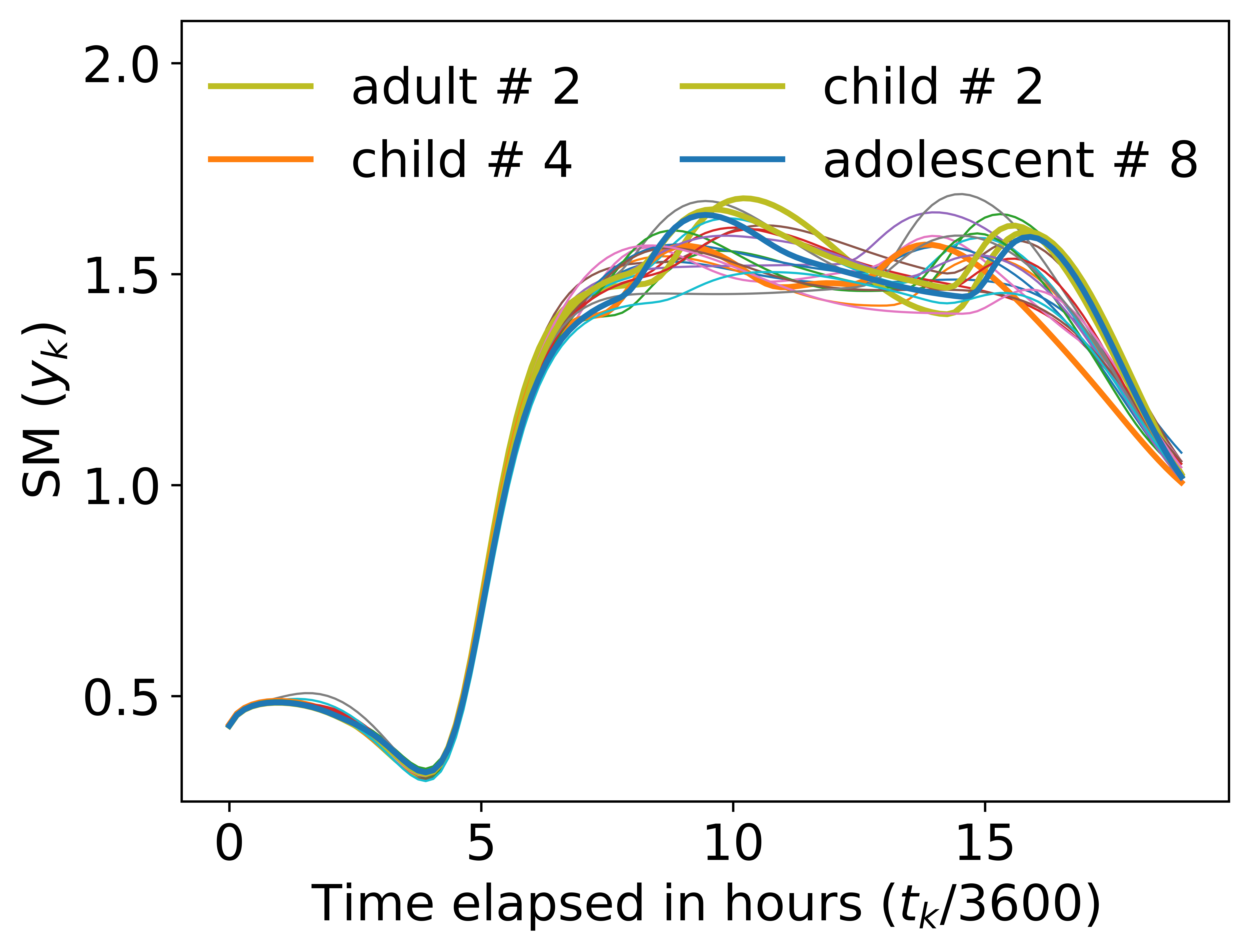}
    \label{fig:figure2}
    }
    \caption{BGL profiles corresponding to the respective 20 virtual patients}
    \label{sensor_ip_op}
\end{wrapfigure}
FDA-approved simulation models of type-I diabetic patients \cite{kovatchev2009silico}, is used to synthetically generate BGL values for 20 virtual patients of diverse age-groups subject to varied food intake and insulin injection profiles over a period of 19 hours (5:00 am to 12:00 am), sampled at every 3 minutes. These 20 virtual patients correspond to  7 adults, 9 adolescents, and 4 children. The BGL profiles of these virtual patients after individually corrupted by white noise of 55 dB signal-to-noise ratio (SNR) are shown in Figure \ref{fig:figure1}, which depicts the variance of BGL profiles in the population. 

\begin{figure}[htbp]
    \centering
    \subfigure[]{
    \includegraphics[width=0.315\textwidth]{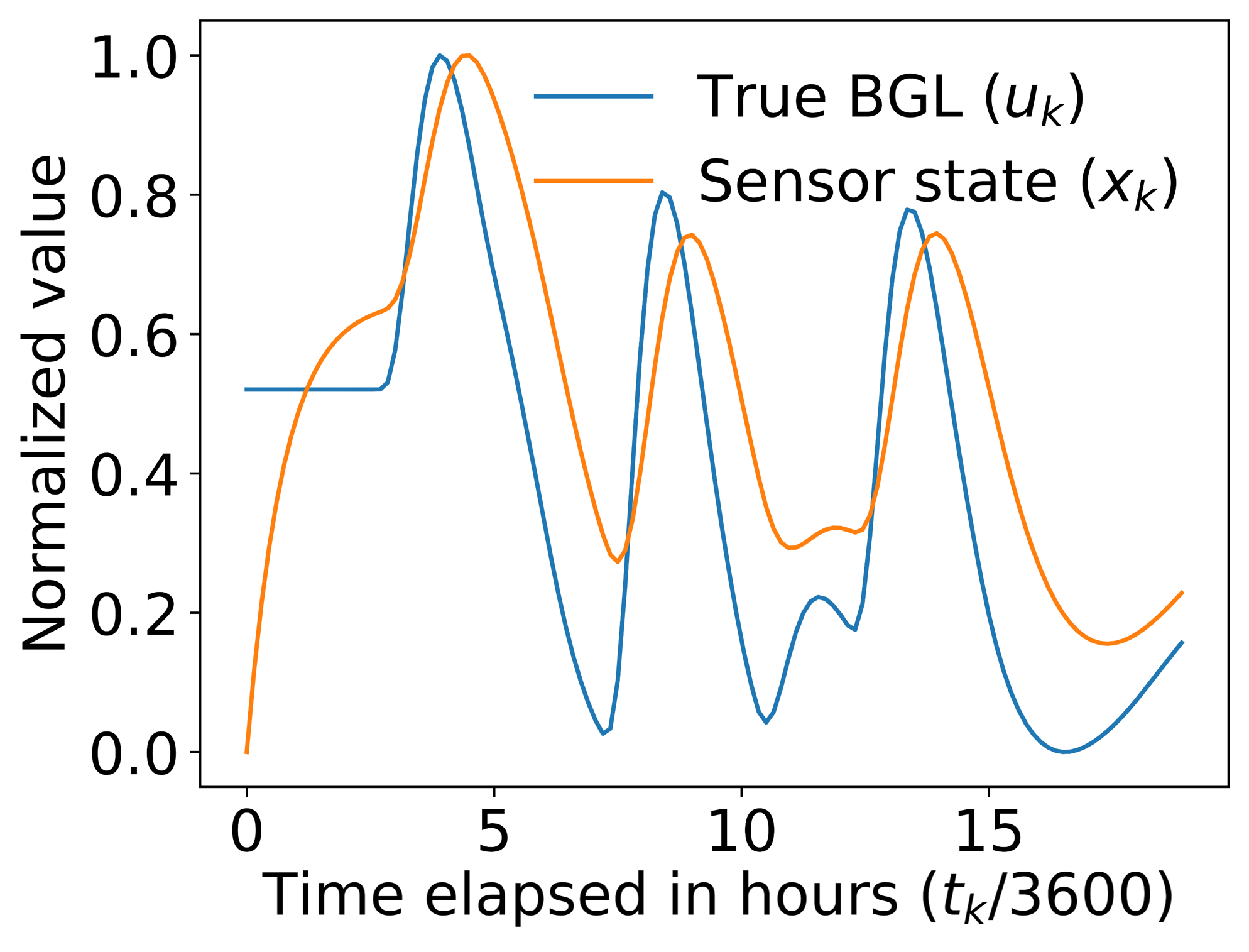}
    \label{fig:figure5}
    }
    \hfill
    \subfigure[]{
    \includegraphics[width=0.315\textwidth]{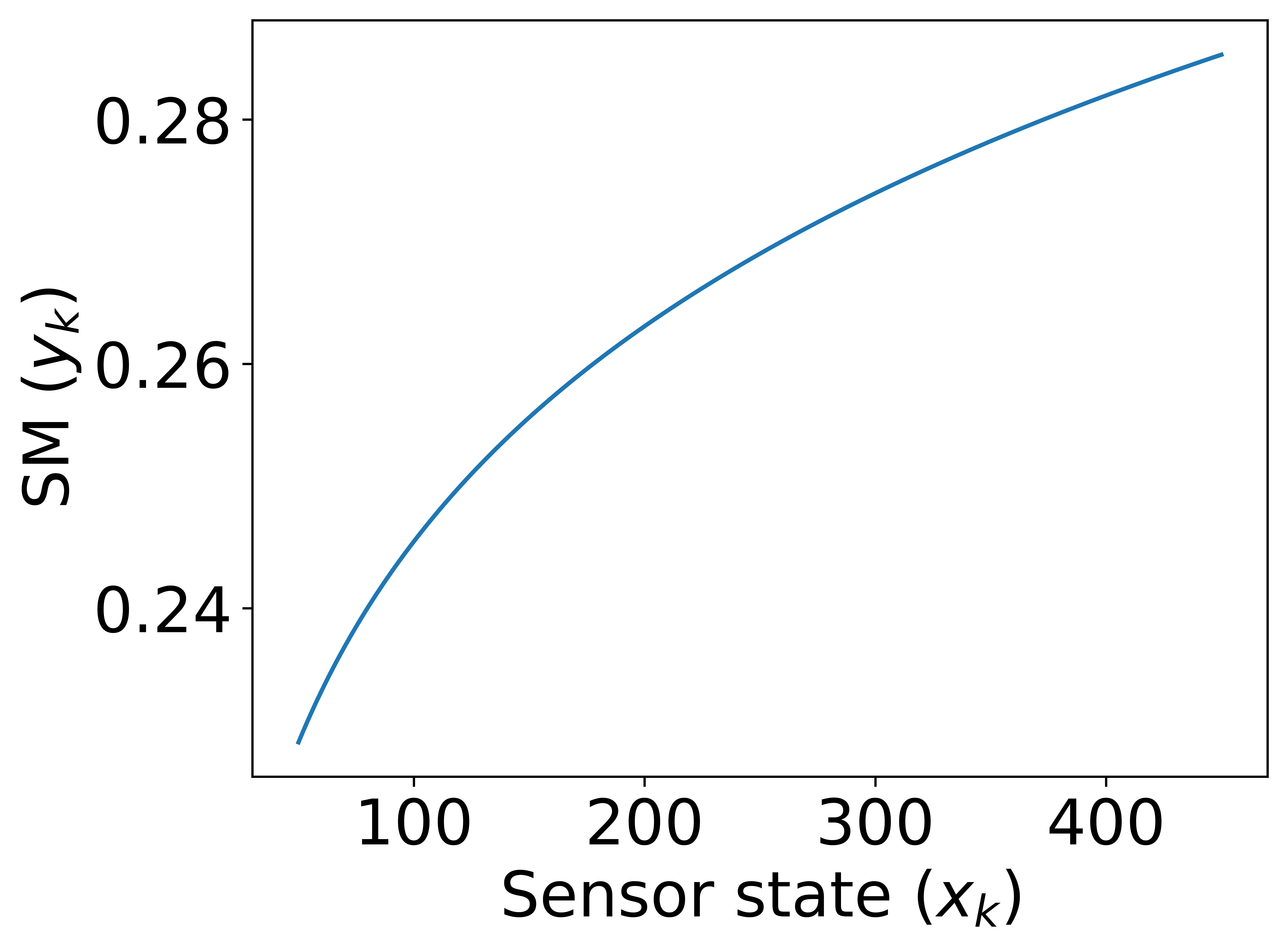}
    \label{fig:figure3}}
    \hfill
    \subfigure[]{
    \includegraphics[width=0.315\textwidth]{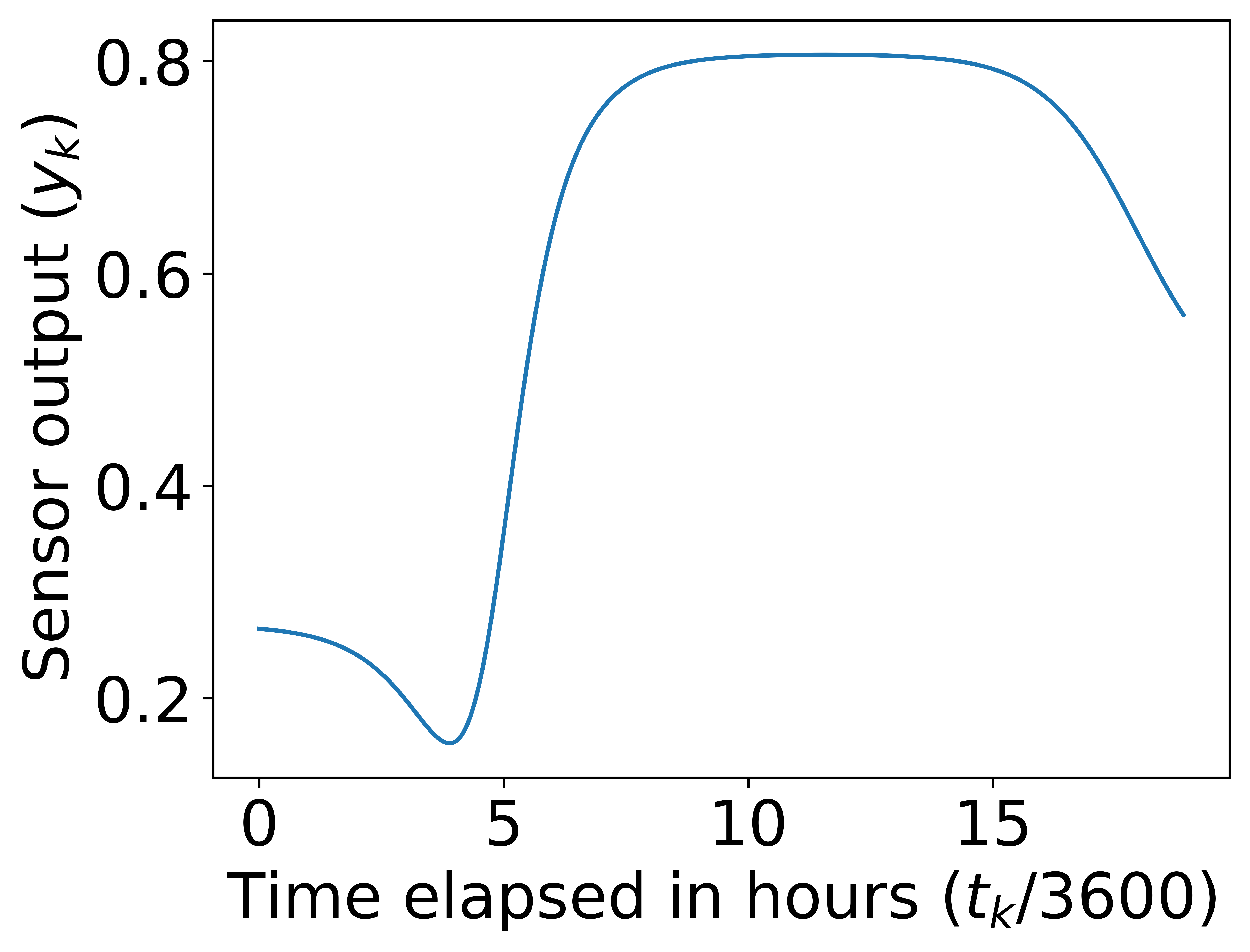}
    \label{fig:figure4}
    }
    \caption{Characteristics of the illustrative sensor model: (a) sensor latent state $x_k$ versus true BGL $u_k$ for adult $\#~2$, (b) Sensor measurement $y_k$ versus state $x_k$ for elapsed time $t_k=50$~min, and (c) Sensor measurement $y_k$ versus elapsed time $t_k$ with state held constant at $x_k=180$.}
    \label{sensor_char}
\end{figure}

A sensor model of the type (\ref{ideal_model}) used in these studies is as follows:
     \begin{align}
         x_{k+1} &= 0.8187x_k + 163.1u_k, \label{dyn_ex}\\
    y_k &= 0.2 (x_k)^{0.1}\Big\{0.8+\tanh(t_k/3T-5).s(t_k/3T-5).\big(0.6+s(17-t_k/3T)\big)\Big\},\label{state-ip-to-op_map_ex}
     \end{align}
where $s(\cdot)$ is the sigmoid function, $T=19\times60=1140$ min, and $t_k$ is measured in seconds. The dynamics of (\ref{dyn_ex}) is simply the zero-order-hold (ZOH) discretization \cite{chakrabortty2015digital} of the continuous-time first-order stable transfer function from $u$ to $x$ with time constant of 15 min (which btw is 5 times the sample interval, thus satisfying the Nyquist's sampling criterion) and DC gain of 900. The normalized plot of $x_k$ versus $u_k$ for the second adult patient model (abbreviated, adult \# 2) is shown in Figure \ref{fig:figure5} revealing the random delay underlying the sensor's characteristics---the dependence of $y_k$ on $x_k$ (resp., elapsed time $t_k$) at a given $t_k$ (resp., $x_k$) is shown in Figure \ref{fig:figure3} (resp., Figure \ref{fig:figure4}), which demonstrates the nonlinear dependence of the output on the state (resp., nonlinear drift of the sensitivity). The above sensor model maps the true BGL profiles of 20 virtual patients to the respective sensor measurement profiles as shown in Figure \ref{fig:figure2}. Figures \ref{fig:figure1}-\ref{fig:figure2} together depict the challenge of decoding the true glucose profiles from the sensor’s measured signal.

Next, given the discrete time-series of the $y_k$'s and the noisy reference BGL values $\tilde{u}_k$'s, our goal is to validate the effectiveness of the proposed SDCM method in computing the estimates $\hat u_k$'s. To accomplish this, we first downsample each time-series of data to the sample interval of 9 min. Next, 16 virtual patients (80\% of the population) are selected randomly to construct the calibration
\begin{wraptable}{r}{2.55in}
\caption{Initial Guess and Range of Parameters}
\label{table1}
\centering
\begin{tabular}{ccc}
\hline
Parameter & Initial Value & Range\\
\hline
$\delta$ & $1$ & $[10^{-5},~10^4]$\\
$\sigma$ & $1$ & $[10^{-5},~10^4]$\\
$\sigma_{\tilde{u}}$ & $10^{-3}$ & $[10^{-5},~10^{-1}]$\\
\hline
\end{tabular}
\end{wraptable}
dataset $\mathcal{D}$, keeping the remaining 4 virtual patient data to test the estimation accuracy. We choose the parameters of the measurement vector $\mathbf{y}_{k,p,\ell}$ to be $p=6,\ell=1$, resulting in a total $N=2040$ training samples and 480 test samples. The overall method is implemented using Scikit-learn 1.0.2 in Python 3.7.9, where the number of trials to find $\theta^*$ is set to 100, and the search for the optimal
value of each parameter in $\theta$ is performed by independently choosing their log-uniform random samples over the respective user-specified ranges, starting from their user-specified initial guesses and evaluating the corresponding likelihood of $\mathcal{D}$ (as in (\ref{likelihood})-(\ref{max_likelihood})). The initial guess and the specified range of the parameter values used for the study are listed in Table \ref{table1}.

\begin{wrapfigure}{r}{0.7\textwidth}
    \centering
    \subfigure[Adult \# 2]{
    \includegraphics[width=0.32\textwidth]{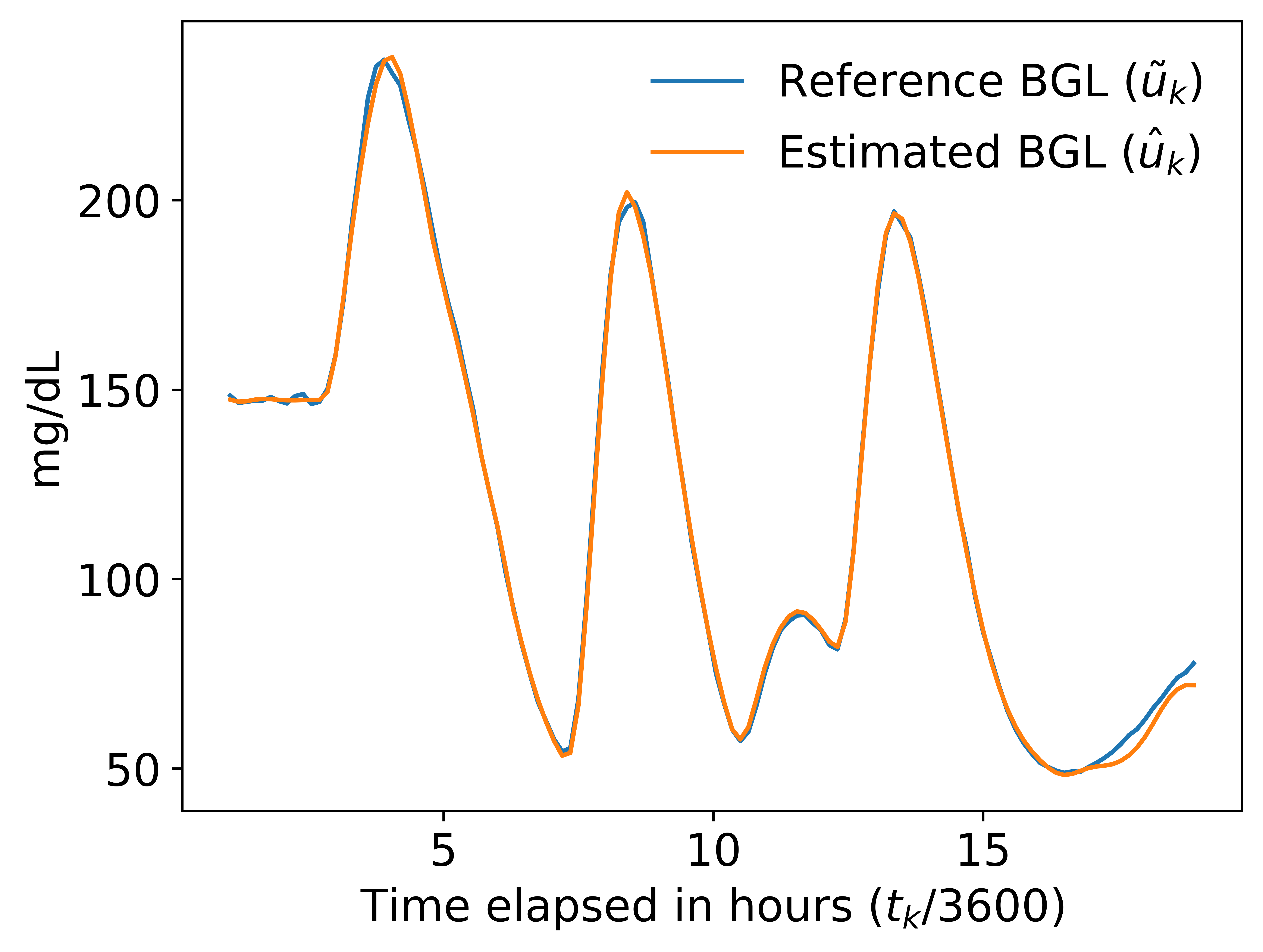}
    \label{fig:figure6}
    }
    \subfigure[Child \# 2]{
    \includegraphics[width=0.32\textwidth]{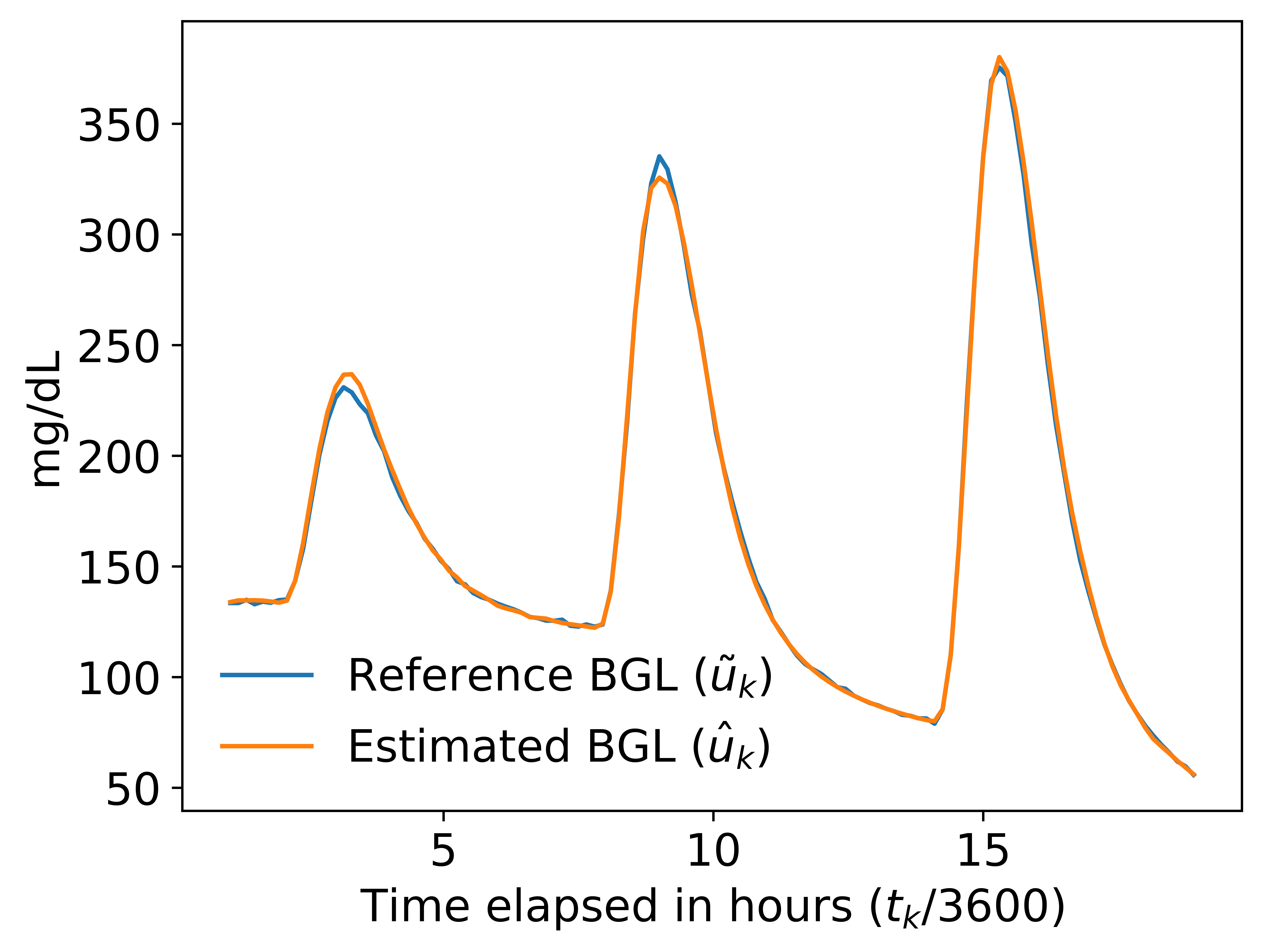}
    \label{fig:figure7}}
    \subfigure[Child \# 4]{
    \includegraphics[width=0.32\textwidth]{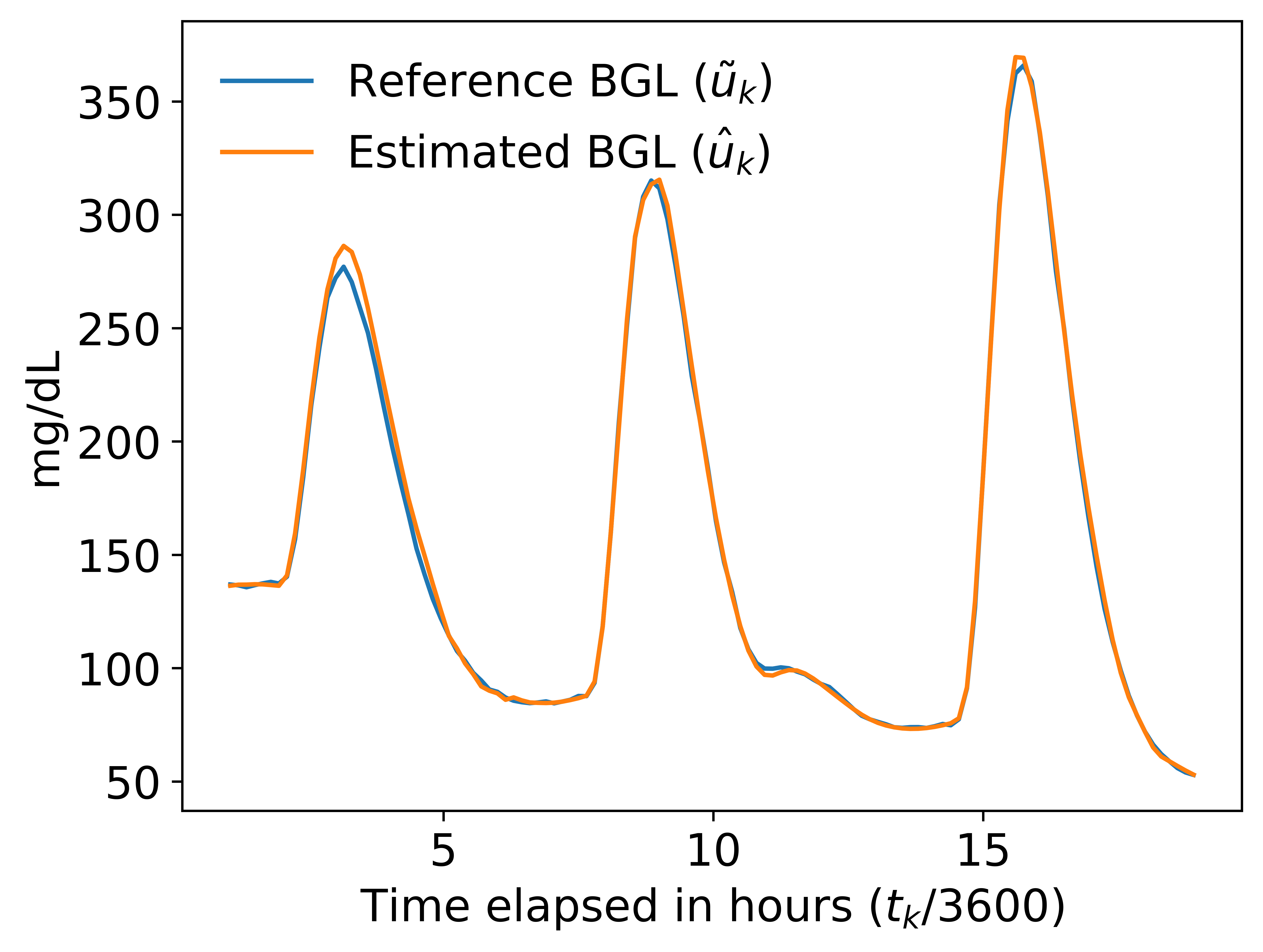}
    \label{fig:figure8}
    }
    \subfigure[Adolescent \# 8]{
    \includegraphics[width=0.32\textwidth]{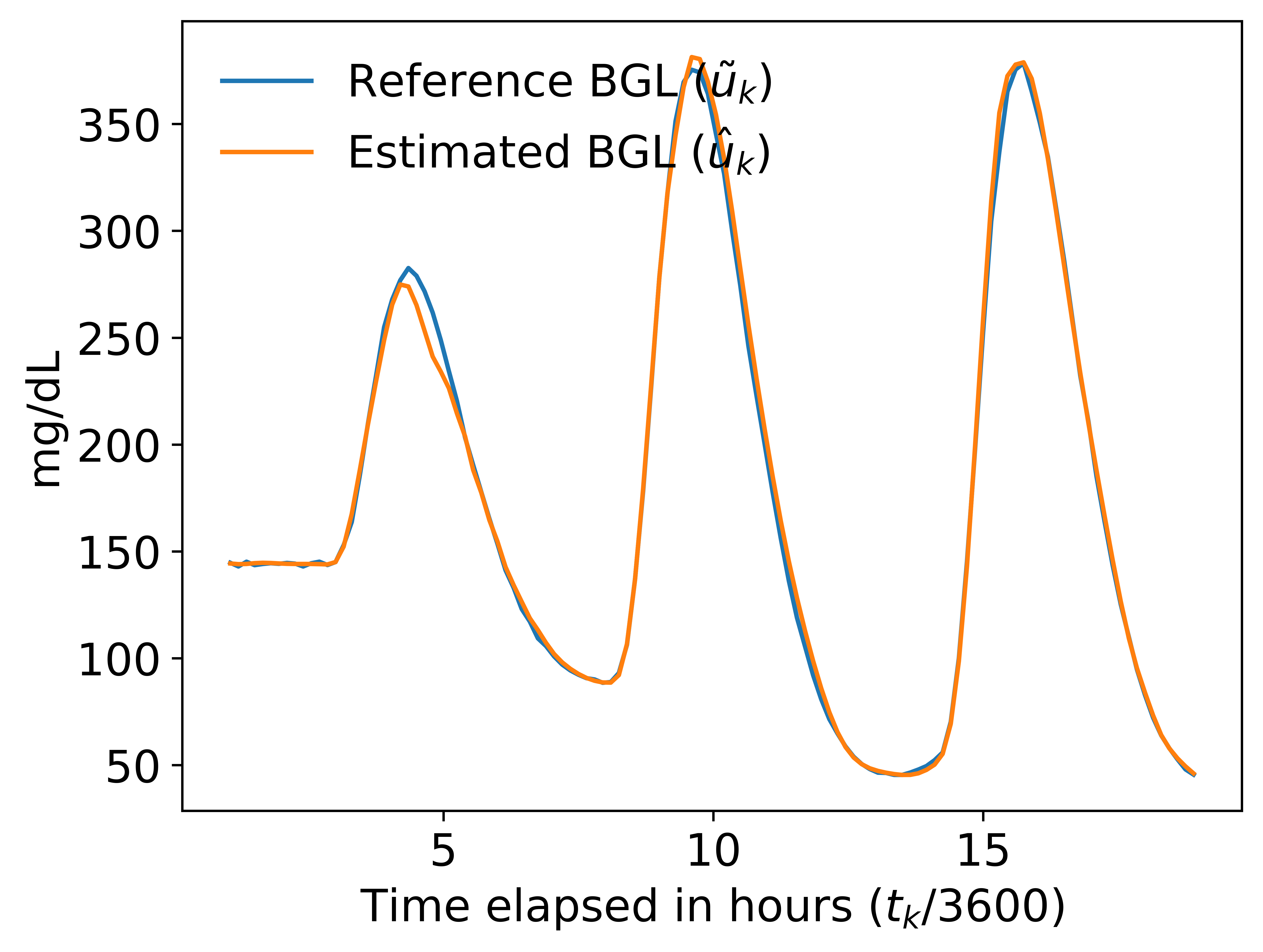}
    \label{fig:figure9}
    }
    \caption{Test performance on a single trial: reference BGL (corrupted by 55 dB white noise) versus estimated BGL}
    \label{individual_plots}
\end{wrapfigure}

\begin{wrapfigure}{r}{2.3in}
    \centering
    \includegraphics[width=2.3in]{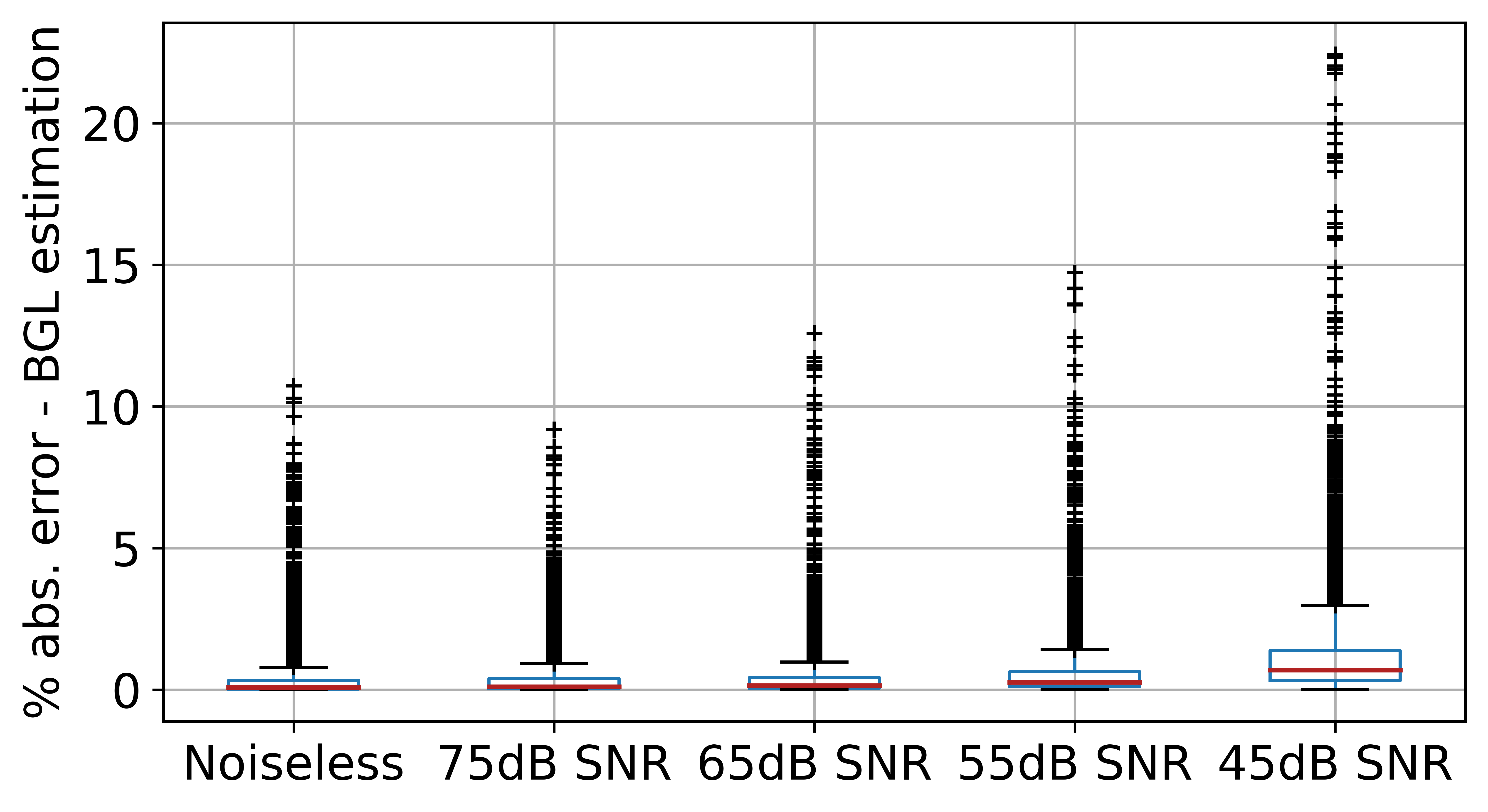}
    \caption{Test error boxplot of 10-fold cross validation and its dependence on the noise level in BGL references}
    \label{noise}
\end{wrapfigure}

The test results for a thus trained SDCM are shown in Figure \ref{individual_plots}, where adult \# 2, child \# 2, child \# 4, 
and adolescent \# 8 are the virtual patients randomly chosen for testing, whereas the remaining 16 virtual patients constitute the calibration dataset $\mathcal{D}$ for training. When compared to the noise ground truth, the estimation error was within 9.83\% of the true BGL uniformly, which is quite encouraging (see explanation towards the end of this paragraph). Next, the sensitivity of the estimation accuracy to the SNR of the reference BGL, along with the robustness to random initial seeds, are verified by performing 10-fold randomized cross-validation trials for a range of SNRs. The boxplots of the percent absolute errors (PAE) with respect to the true BGLs of all test samples of the 10 random trials under different SNR levels is shown in Figure \ref{noise}. Note the outliers in the boxplots correspond to the predictions lying {\em roughly} beyond 2 times the inter-quartile range (IQR) of all PAEs on either side of the median value (the term ``roughly" is used to qualify
the fact that there is no a priori guarantee of symmetry of IQR around the median). As seen in the figure, lower SNR of the reference BGL results in a larger mean and variance of the absolute error percentage (i.e., inferior estimation quality). Yet the estimation accuracy of the models trained using reference values of < 55 dB SNR, which is typical for calibrations in practice, comfortably satisfies the FDA ISO 15197 2013 standard that requires the error of 95 \% of the estimated samples to be within 15\% in case of true BGL > 100 mg/dL, and to be within 15 mg/dL otherwise \cite{klonoff2015performance}.  Including both training and evaluation, each scenario of a given SNR of reference BGL took approximately 25 minutes on a standard laptop with 64-bit operating system, 16 GB RAM, and a 2.6 GHz processor.

As mentioned earlier, the existing methods of \cite{vettoretti2015online, vettoretti2019development} that also learn temporally drifting calibration maps for CGM applications are restricted to be linear and second-order polynomial, respectively, and
in case of high nonlinearity in the sensor's temporal drift as is the case in Figure \ref{fig:figure4}, both these methods will clearly result in much larger errors compared to ours. Calibration maps using higher degree polynomials may be capable of approximating the nonlinearity present in the real-life data but with a high risk of over-fitting given a limited-sized dataset. Additionally, unlike \cite{vettoretti2015online, vettoretti2019development}, our method does not require a knowledge of the underlying dynamics of the sensing system and hence is more generally applicable. 

\begin{wrapfigure}{r}{2in}
    \centering
    \includegraphics[width=1.5in]{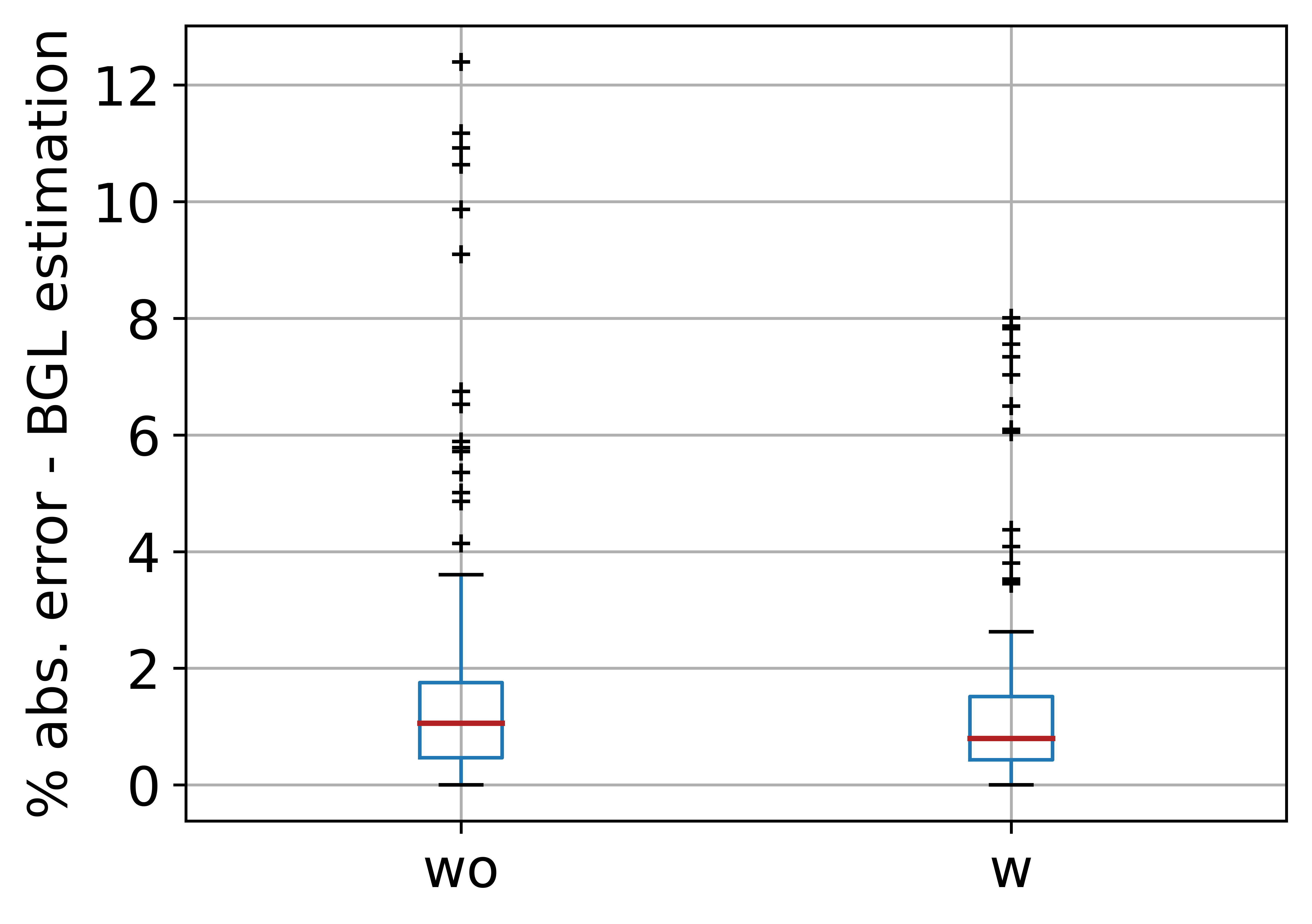}
    \caption{Test PAE boxplots of two cases: with (w) and without (wo) online iterative update}
    \label{update_perf}
\end{wrapfigure}

To validate the performance of Algorithm \ref{algo1}, we tune the parameters $(\epsilon_u,c,\epsilon_\gamma)$ empirically by latin 
hypercube sampling \cite{mckay2000comparison} over $[0,0.1]\times[0.2,1.0]\times[2.0,7.0]$ with each range divided into 6 uniform-length subintervals. The randomly sampled triple $(0.0684,0.7346,6.3445)$ is chosen as optimal since updating the calibration dataset and retraining an above-trained SDCM employing Algorithm \ref{algo1} along with these parametric values, using the first 9 hours of data corresponding to an independent patient model, results in prediction with the least mean PAE on the last 10 hours of that data. Using the optimal $(\epsilon_u,c,\epsilon_\gamma)$, we rerun the process on a second independent patient model and compare the prediction performance over the last 10 hours with the case of no online update. The boxplots comparing these two cases, as shown in Figure \ref{update_perf}, depict the usefulness of Algorithm \ref{algo1}. 

 \subsection{Calibration of experimental data from Real Continuous Glucose Monitoring Data}
To validate our method on real-life CGM data, we assembled an optical glucose sensor at our laboratory, which is used to conduct CGM experiments on alive rats. A description of the working
principle of the sensor and the arrangement adopted for CGM on a rat subject are provided in Appendix \ref{sensor_desc} and \ref{rat_desc} respectively. During the assay,
rat blood samples were collected every 3 min to obtain the reference BGL values (i.e., the $\tilde{u}_k$'s) for calibration using a commercial handheld glucometer, and simultaneously, also the optical sensor reflectance values at its resonant wavelength (i.e., the $y_k$'s) were recorded. It was observed that the deposition of blood constituents on the surface of sensor's microdialysis probe membrane causes drift of the sensor's sensitivity over time, and it also limits the life of the sensor. In our study, we found that a probe remains usable for about 120 min, sufficient to perform the experiment. We obtained the experimental data for four such 120 min periods, employing a new probe in each period.

For the calibration, the parameters of the measurement vector $\mathbf{y}_{k,p,\ell}$ was set to $p=8,\ell=1$. Accordingly, a dataset comprising $N=124$ data samples of the form $(\overline{\mathbf{z}}_k, \tilde{u}_k)$ was generated. The choice of the initial 
guesses, the ranges of the respective parameters in $\theta$, and the method of
\begin{wrapfigure}{r}{2in}
    \centering
    \includegraphics[width=1.5in]{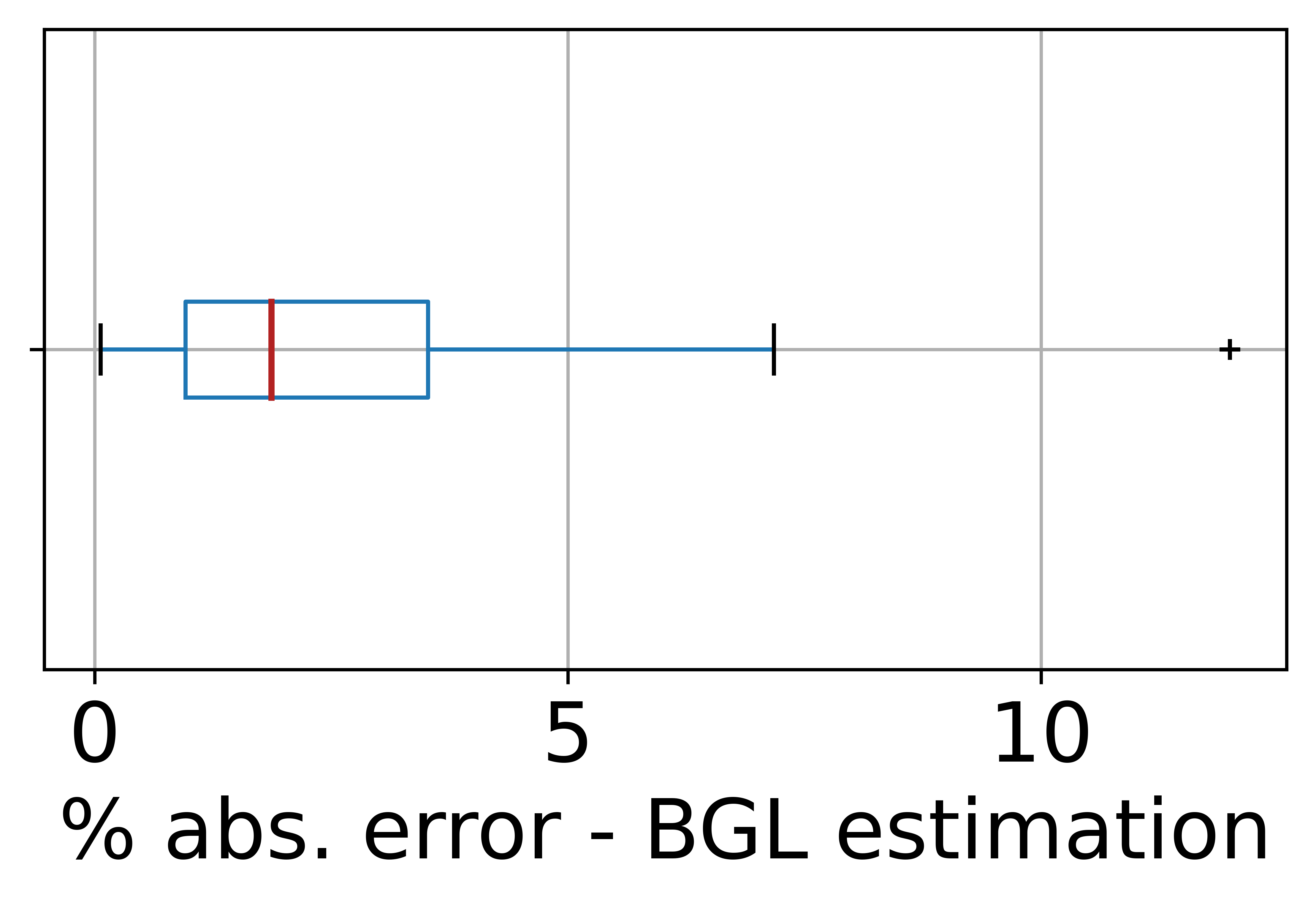}
    \caption{Test PAE boxplots of 10-fold cross validation on CGM data}
    \label{real_data_perf}
\end{wrapfigure}
tuning those were the same as used in Section \ref{synthetic}. Owing to the limited size of the available dataset, we adopted a 10-fold randomized cross-validation strategy on the total dataset without segregating them into the four CGM periods. The dataset was randomly divided in 85:15 proportion for training versus testing respectively, and the process was repeated 10 times independently. The percent BGL estimation errors were computed with respect to the corresponding ground truth values. The boxplot of the PAE of all test samples in the 10 random trials is shown in Figure \ref{real_data_perf}. The maximum estimation error was found to be $11.92\%$, which again meets the requirements of FDA ISO 15197 2013 standard \cite{klonoff2015performance} and is also comparable to the case of synthetic noisy data discussed in the previous subsection. 

\section{Conclusion}
To achieve flexibility in designing sensors by allowing nonlinearity, time-drift, and random delay in its underlying dynamics, a novel Bayesian inference-based dynamic calibration method is proposed, where the sensed value is estimated as a posterior conditional mean given a finite sequence of the sensor measurements and the elapsed time. In addition, an iterative algorithm is proposed for online update of the estimated value upon the arrival of a new data, where the newly collected data is first evaluated for not being an outlier, and when so, it is use to replace an appropriately chosen existing data. The method's effectiveness is validated on CGM data from an alive rat using an optical glucose sensor developed in our lab. To allow flexibility in choice, the validation is also performed on a synthetic blood glucose level dataset generated using FDA-approved virtual diabetic patient models coupled with an illustrative CGM sensor model. The work reports for a first time a dynamic calibration approach for sensors with nonlinearity, time-drifts, and random delays, and validates it over real-life measured data for glucose monitoring.


\medskip

\small
\bibliography{References}
\bibliographystyle{plainnat}

\appendix
\section{Appendix}
\subsection{Optical Glucose Sensor and Measurement Setup}\label{sensor_desc}
We fabricated a gold nanoparticles (AuNP)-coated optical fiber sensor probe for in-vivo glucose detection in blood. In-vivo detection requires a reflected signal for measurement. Hence the sensor is made at the distal end of the fused optical fiber. The light guided through the optical fiber interacts with the AuNPs at the distal end of the fiber and excites the local surface plasmon resonance (LSPR), where at the resonance wavelength, the reflected light has maximum absorption and minimum reflected intensity. The resonance wavelength depends on AuNP characteristics---material, size, shape, and refractive index (RI)---and RI of the local surrounding medium. A change in the local surrounding medium RI results in a shift in resonance wavelength as well as a change in the absorption level (so, the intensity level of reflected light) at the original resonant wavelength. For specificity, the AuNP-coated fiber is functionalized with the glucose oxidase (GOx) enzyme, which specifically oxidizes glucose molecules and thereby exhibits shifts in the reflected light intensity at LSPR in response to different glucose concentrations.

\begin{figure}[h!]
\centering\includegraphics[width=\linewidth]{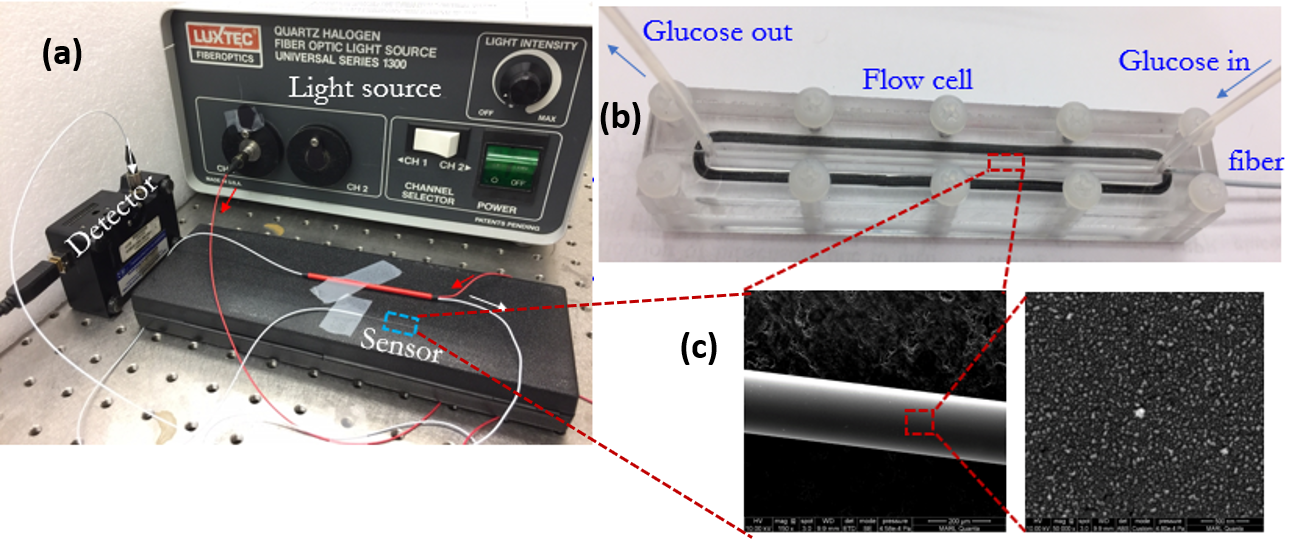}
\caption{Experimental setup of Fiber Optic Sensor}\label{fig:ISU_setup}
\end{figure}

Fig. \ref{fig:ISU_setup} depicts the optical sensor's experimental setup, consisting of a broadband white light source, an optical detector (range 400~nm to 750~nm), and a 2x1 fused optical fiber. The light source is used to illuminate the sensing element of approximately 4 cm length toward the distal end of the optical fiber. Light reflected off the element is detected by the detector. The scanning electron microscope image in the inset of Fig. \ref{fig:ISU_setup} confirms the successful attachment (using (3-aminopropyl)trimethoxysilane linkers) of AuNPs to the fiber surface, which leads to local surface plasmon resonance (LSPR). The response of the sensor to glucose was independently validated in the lab before in-vivo use by employing a flow cell around the sensing element through which distilled water-based glucose solutions of various glucose concentrations (within  0-450 mg/dL range) are passed, and the corresponding spectra of reflectance values of the sensing element are observed.

\subsection{Continuous Glucose Monitoring Setup on Rat and Experimental Plots}\label{rat_desc}
\begin{figure}[h!]
\vspace*{-.15in}
\centering\includegraphics[width=0.8\linewidth]{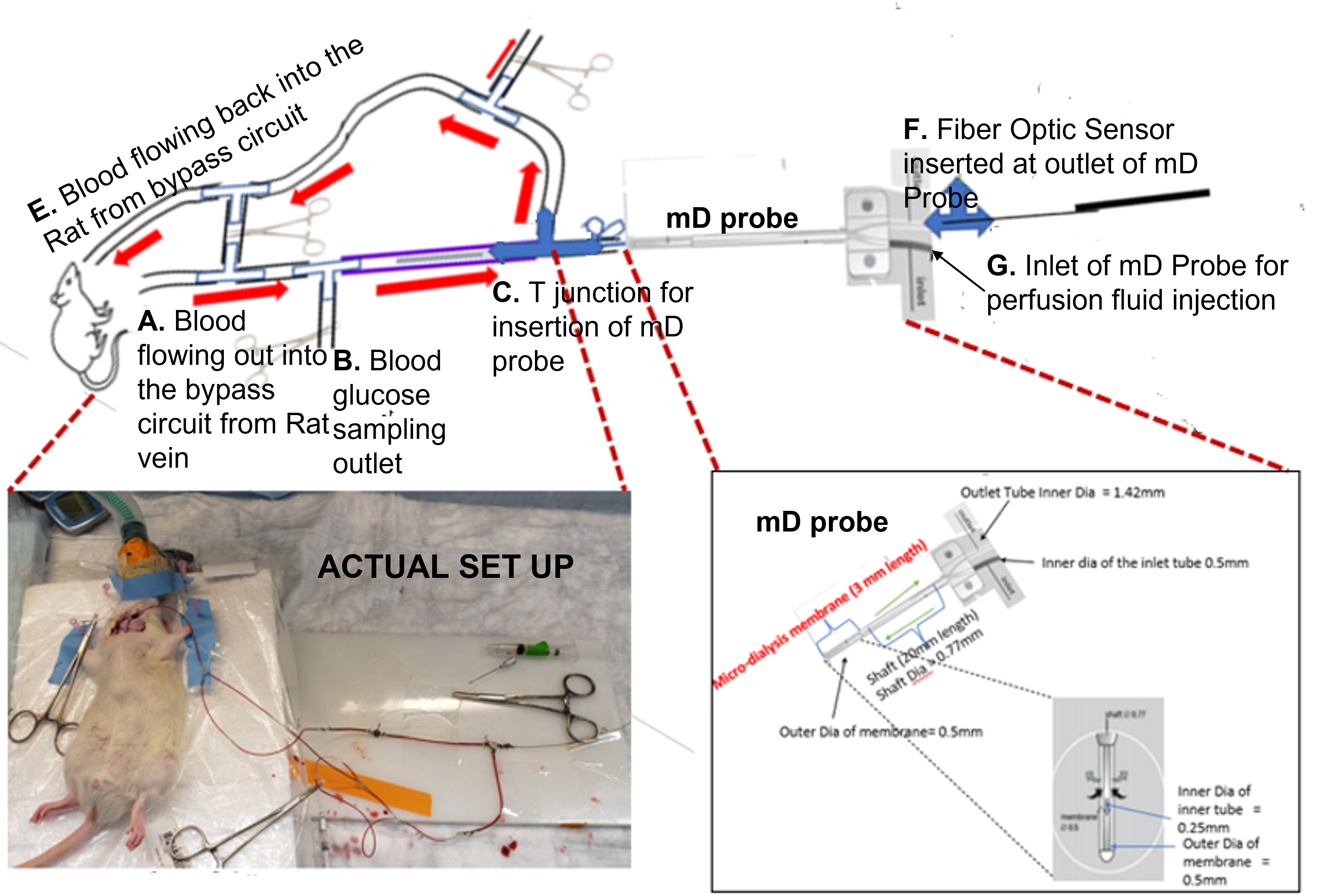}
\caption{Experimental Setup for CGM on alive rat}\label{fig:UoI_Setup}
\end{figure}

Figure \ref{fig:UoI_Setup} shows the experimental setup for CGM on an alive rat, along with the schematics of a bypass circuit and a microdialysis (mD) probe. To prevent the optical sensor from fouling from sample deposition and coagulation, thereby extending its life and improving its accuracy, the sensor was inserted at the outlet of the mD probe (MWCO 20kDa, chosen in accordance with the molecular weight of glucose). There are seven main ``points" of the experimental setup as shown alphabetically in the schematic diagram of Figure \ref{fig:UoI_Setup}. Point A depicts blood flow from the rat's right jugular vein into the surgically set bypass circuit. Point B is periodically opened and closed to sample blood from the bypass circuit to obtain the reference values of blood glucose concentration using a commercial glucometer. At Point C, a 3-way T-junction is placed to reroute the blood back into the rat, and also the mD probe is placed in the junction as shown in the figure to allow any blood components less than 20kDa to pass through the mD membrane, blocking the larger ones. Point D is an inlet made available for occasional washing of the external surface of the mD to remove any blood debris clinging onto the probe surface. At point E, the blood flows back into the jugular vein of the rat, thereby closing the bypass loop. The mD probe has one inlet (Point G) and one outlet (Point F). The inlet Point G is fed with heparin as a perfusion fluid, which prevents blood from clotting at the interface of the mD membrane, through which the glucose molecules percolate into the mD probe.
The LSPR fiber optic sensor is placed at the outlet Point F to sense the incoming analyte solution containing glucose. The BGL is adjusted in the rat by managing the insulin and the glucose infusion rates via the right jugular catheter. For a normal rat with insulin flow at 2.5 mU/min/kg, we observed that a typical glucose infusion rate of about 55-60 mg/kg/min is needed to maintain the BGL steady at its initial value.

\begin{figure}[h!]
\vspace*{-.2in}
\centering\includegraphics[width=0.4\linewidth]{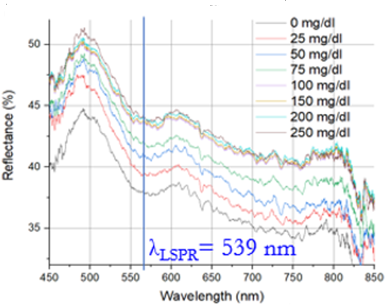}
\caption{Reflectance data from LSPR Sensor (Resonant wavelength 539nm)}\label{fig:LSPR_Signal}
\end{figure}

Fig.\ref{fig:LSPR_Signal} depicts the reflectance spectra of the LSPR sensor measurement for different glucose concentration from 0~mg/dl to 250~mg/dl. The resonant wavelength is at 539~nm where the reflected light intensity minima was observed.  

In the lab measurement setting, a continuous shift in the reflectance versus the glucose concentration provided a measure for the calibration of our fiber-optic glucose sensor as explained in Section 4.2.

\end{document}